\documentclass[12pt, reqno]{article}
\usepackage{graphicx, amsmath, amssymb, amsthm, amsfonts, cite, hyperref, epstopdf}
\usepackage[lmargin=2.25cm, rmargin=2.25cm, tmargin=2.5cm, bmargin=3cm]{geometry}
\newcommand{\ka}{\kappa}
\newcommand{\ed}{\end{document}}
\newcommand{\be}{\begin{equation}}
\newcommand{\ee}{\end{equation}}

\begin{document}

%\maketitle

\title{\textbf{Ideal Gas Thermodynamics with an Invariant Energy Scale}}%
\author{Dr. Sudipta Das \\ Assistant Professor \\ Department of Mathematics \\ Banwarilal Bhalotia College, Asansol - 713303, India \\ E-mail: sudipta$\_$jumaths@yahoo.co.in}%

\maketitle

%\address{}%
%\email{sudipta$\_$jumaths@yahoo.co.in}%

%\thanks{}%
%\subjclass{}%
%\keywords{Ramanujan Primes}%

%\date{}%
%\dedicatory{}%
%\commby{}%
% ----------------------------------------------------------------
\vspace{.8cm}

\begin{abstract}
  A viable approach towards Quantum Gravity is the Doubly Special Relativity (DSR) framework in which an observer-independent finite energy upper bound (or a finite smallest length scale) appears quite naturally. In this work, we have studied the thermodynamic properties of an ideal gas in a specific DSR framework, known as the Magueijo-Smolin (MS) Model. We use the fact that DSR can be considered as nonlinear representation of Lorentz Group. Subsequently, various thermodynamic parameters of ideal gas have been derived in this modified framework to compare the corresponding deviations from the usual scenario due to the presence of the invariant energy (length) scale.
\end{abstract} 

\vspace{1.2cm}

\section
{Introduction}

One of the most important as well as challenging problem in Modern Theoretical Physics is to construct a theory of Quantum Gravity. Although many promising approaches have been proposed, a final theory which can successfully merge Quantum Field Theory with Theory of Gravitation still eludes us. As there is a smallest length scale (Planck length) or largest Energy scale (Planck energy) inherent in Quantum Mechanics, any theory of Quantum Gravity is also expected to depict such behaviour. According to the Relativity Principle, these length or energy scales should be observer-independent. However, this is in direct contradiction with Einstein's Special Relativity (SR) as the Lorentz transformation laws make the quantities like length and energy (or mass) of an object to be observer-dependent. Thus, a modification to the Theory of Special Relativity is required to incorporate any such observer-independent quantities. One such modified theory is Doubly Special Relativity (DSR), first proposed by Amelino-Camelia \cite{ac1}. Subsequently, other simpler models of DSR have been introduced by Amelino-Camelia \cite{ac2}, Magueijo and Smolin \cite{mag1}. The crucial aspect of all these theories of DSR is the presence of another observer-independent invariant quantity, an energy upper limit denoted by $\ka$, in addition to the constant velocity of light in vacuum, $c$. To make this invariant energy (or length) scale compatible with SR, the usual relation between energy and momentum of a particle, known as the dispersion relation (or mass-shell condition) has to be modified in DSR. As a result, the transformation laws in DSR become nonlinear through modifications of linear Lorentz transformations of SR (please see the reviews \cite{acr, kowglik1} and references therein for detail discussion on DSR). \\
Various thermodynamic properties of Photon gas using a particular dispersion relation have been thoroughly discussed in \cite{camacho1, zhang}. In \cite{magcos, bertolami}, the authors have studied thermodynamics of Boson and Fermion gases and their astrophysical properties by considering another dispersion relation. In this work, the Magueijo-Smolin (MS) dispersion relation \cite{mag1} has been considered. The expression for Energy-momentum tensor of a perfect fluid in this MS model has been derived \cite{dasghosh} and thermodynamics of Photon gas with the same MS dispersion relation has also been studied \cite{daschow}. In this article, we discuss the thermodynamic properties of ideal gas in non-relativistic limit of MS model where deviations from the usual results due to the presence of an energy upper bound have been analyzed.\\
The paper is organized as the following: in the next Section 2 after Introduction, the methodology for this work has been described. Starting with the MS dispersion relation, the partition function for ideal gas in non-relativistic limit has been derived. This is the central point of this whole work. In the following Section 3, different thermodynamic parameters like free energy, entropy and others have been calculated using this expression for partition function. Furthermore, these results have been plotted side by side with the usual scenario to compare the deviations due to the invariant energy scale. In the final Section 4, we summarize and conclude our article with remarks regarding future prospects.

% ----------------------------------------------------------------

\section{Modified Dispersion Relation and Partition Function for Non-relativistic Ideal Gas}

As discussed in the previous Section, the usual dispersion relation for a particle in Special Relativity (SR) is given by 
\be E^2 - p^2 = m^2, \label{dr} \ee where $E$ is the energy, $p$ is the three-momentum magnitude and $m$ is the rest mass of the particle and the velocity of light in vacuum has been considered to be unity, $c = 1$ to make the dimensions of energy and momentum to be uniform. In Doubly Special Relativity (DSR), this dispersion relation (\ref{dr}) is modified with introduction of an observer-independent invariant energy scale (or length scale). Various models of DSR have been proposed with different forms of modified dispersion relations \cite{ac1, ac2, mag1}. In this article, we have considered the Magueijo-Smolin (MS) model \cite{mag1}, where the modified dispersion relation is expressed as
\be E^2 - p^2 = m^2 \left(1 - \frac{E}{\ka}\right)^2, \label{mdr} \ee $\ka$ being the invariant energy upper bound. It can be clearly seen that (\ref{mdr}) reduces to (\ref{dr}) in the limit $\ka \rightarrow \infty $. As opposed to the other dispersion relations as considered in \cite{camacho1, zhang, magcos, bertolami}, this MS dispersion relation (\ref{mdr}) is known to have more theoritical motivations. In \cite{sghosh}, it was shown that the energy (or length) scale $\ka$ in (\ref{mdr}) is theoritically consistent with a non-commutative symplectic phase space structure, named as $\ka$-Minkowski spacetime, by modifying the usual Poisson brackets of position and momentum. This leads to modification of the usual linear Lorentz transformations to nonlinear $\ka$-Lorentz transformations under which the MS dispersion relation (\ref{mdr}) remains covariant \cite{bruno, sghosh}. In \cite{hoss}, it has been further shown that in this MS model \cite{mag1, dasghosh, daschow}, the velocity of light $ c = \frac{d	E}{d p} $ truly remains an invariant quantity whereas it becomes energy-dependent in case of the other dispersion relations as considered in \cite{camacho1, zhang, magcos, bertolami}. \\
In this work, our primary task is to derive an expression for the partition function of an ideal gas for MS model in DSR framework. For a non-relativistic ideal gas, the energy $ E $ or Hamiltonian $ H (p, q) $ of a single particle is simply given by 
\be E = H (p, q) = \frac{p^2}{2m}, \label{nrh} \ee 
where $p$ is the magnitude of three-momentum, $q$ is the generalized coordinate and $m$ is the mass of the particle. Using (\ref{nrh}), the corresponding single particle partition function $ Z (T, V, 1) $ can be obtained by following the definition \cite{pathria, greiner}
\be Z (T, V, 1) = \frac{1}{h^3} \int e^{- \beta H} d^3 q ~ d^3 p = \frac{4 \pi V}{h^3} \int_0^{\infty} p^2 dp = V \left(\frac{2 \pi m k_B T}{h^2}\right)^{3/2}, \label{nrz1} \ee where $T$ is the temperature of the gas, 
$V = \int d^3 q$ is the volume occupied by the gas, $h$ is the Planck constant and $ \beta = \frac{1}{k_B T} $, $k_B$ being the Boltzmann constant. Considering the ideal gas to be a system of $N$ number of non-interacting particles, one can easily calculate the total partition function $ Z(T, V, N) $ by using the Gibb's factor $ Z(T, V, N) = \frac{1}{N!} (Z(T, V, 1))^N $ and standard integral formulae \cite{pathria, greiner, gradstein} as the following
\be Z(T, V, N) = \frac{V^N}{N!} \left(\frac{2 \pi m k_B T}{h^2} \right)^{3N / 2}. \label{nrzn} \ee
For a relativistic ideal gas consisting of $N$ number of non-interacting particles, the same procedure can be applied to find the expression of the partition function. However, one has to keep in mind that the relativistic Hamiltonian is obtained from (\ref{dr})
\be H = m \left(1 + \left(\frac{p}{m}\right)^2 \right)^{1/2}, \label{rh} \ee which is evidently different from (\ref{nrh}).
Proceeding in similar manner, using this Hamiltonian (\ref{rh}) and doing suitable substititutions, we can evaluate the integral and finally obtain the expression for the total partition function \cite{greiner}
\be Z(T, V, N) = \frac{1}{N!} \left[4 \pi V \left(\frac{m}{h}\right)^3 e^{\beta m} \frac{K_2(\beta m)}{\beta m}\right]^N, \label{rzn} \ee where $K_2$ represents the modified Bessel function. Using the series expansion and differential properties of modified Bessel function, it can checked that (\ref{rzn}) reduces to (\ref{nrzn}) in the non-relativistic limit $ \beta m \rightarrow \infty $, i.e. $ m \gg k_B T $ \cite{greiner}. \\
In this article, we are interested to study the effects of the invariant energy upper bound in case of an ideal gas at a low-energy non-relativistic regime. At first, we proceed to find out an expression for partition function of an ideal gas with the MS dispersion relation (\ref{mdr}). It has been been clearly discussed in the previous Sections that there remains an energy upper bound $\ka$ in this MS model. Thus, the integrals to calculate the partition function have an upper limit $\ka$ whereas the integral (\ref{nrz1}) to derive (\ref{nrzn}), (\ref{rzn}) have been evaluated using infinite upper bounds of energy or momentum. The expression for partition function for classical Photon gas with MS dispersion relation (\ref{mdr}) has been derived in \cite{daschow}. As we have stated earlier, in this particular work we first try to find out the partition function for a classical ideal gas in non-relativistic limit using the same MS model (\ref{mdr}). It is also noteworthy that an analytical expression for partition function of relativistic ideal gas in MS model could not be derived due the presence of the upper limit $\ka$ along with the complicated dispersion relation (\ref{mdr}), which can be studied numerically in future works. For now, we start with the non-relativistic Hamiltonian (\ref{nrh}) and proceed to calculate the single particle partition function of ideal gas in MS model as following \cite{daschow}
\be Z_{MS}(T, V, 1) = \frac{1}{h^3} \int e^{- \beta H} d^3 q ~ d^3 p = \frac{4 \pi V}{h^3} \int_0^{\ka} p^2 e^{-\beta \frac{p^2}{2m}} d p. \label{msz1} \ee
The invariance of the phase space volume $ \int d^3 q ~ d^3 p $ in DSR under $\ka$-Lorentz transformations has been discussed in \cite{daschow}. The upper limit in the integral (\ref{msz1}) can be calculated in terms of error function 
$ Erf(x) = \frac{2}{\sqrt{\pi}} \int_0^x e^{- t^2} dt $ and the final expression for single particle partition function becomes
\be Z_{MS}(T, V, 1) = \frac{4 \pi V}{h^3} \left[\sqrt{\frac{\pi}{2}} (m k_B T)^{3/2} ~ Erf \left(\frac{\ka}{\sqrt{2 m k_B T}}\right) - m k_B T \ka e^{- \frac{\ka^2}{2 m k_B T}} \right]. \label{msz1f} \ee
In the following Figure 1, we have shown the interesting plot of the single particle partition function for both the normal case (\ref{nrz1}) and for MS model (\ref{msz1f}) in case of a non-relativistic ideal gas, drawn side by side. It is evident from the Figure 1 that the partition function for the MS model always stays lower than the usual case and deviates significantly from the normal plot. This is clearly an effect of the energy upper bound $\ka$ present in the MS model of DSR as the number of available microstates for the system becomes finite, which we have discussed later in detail. 
\begin{figure}[htb]
	{\centerline{\includegraphics[width=10cm, height=6cm] {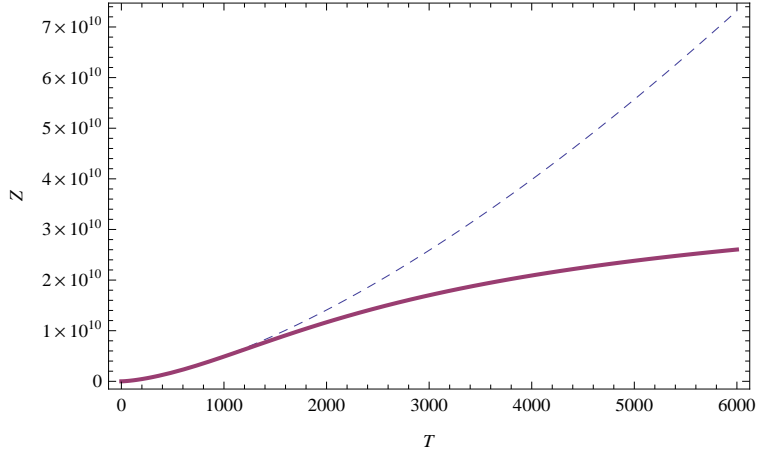}}}
	\caption{{\it{Plot of single particle partition function of non-relativistic ideal gas $S$ against temperature $T$ for both in the usual theory and in MS model with an invariant energy scale. The dashed line corresponds to the usual result and the thick line represents the corresponding quantity in MS model considered in our work. For the sake of convenience, we have scaled the units and the corresponding parameters take the following values $\ka = 10000, k_B = 1, N = 10000, m = 10000, V = .01, h = 1 $ in this plot as well as in all other plots in the paper. In this scale, $T=10000$ is the Planck temperature.}}} \label{fig1}
\end{figure} \\
Applying the Gibb's factor, the total partition function for non-relativistic ideal gas in MS model can be obtained as
\be Z_{MS}(T, V, N) = \frac{1}{N!} \left(\frac{4 \pi V}{h^3}\right)^N \left[\sqrt{\frac{\pi}{2}} (m k_B T)^{3/2} ~ Erf \left(\frac{\ka}{\sqrt{2 m k_B T}}\right) - m k_B T \ka e^{- \frac{\ka^2}{2 m k_B T}} \right]^N. \label{msznf} \ee
It can be easily verified that in the limit $ \ka \rightarrow \infty $, the partition function in (\ref{msznf}) can be simplified to the expression as given in (\ref{nrzn}).
% ----------------------------------------------------------------

\section{Thermodynamic Properties of Non-relativistic Ideal Gas}

Having the expression for partition function in hand (\ref{msznf}), we now proceed further to study thermodynamic properties of non-relativistic ideal gas in MS model to observe the effects of the energy upper bound $\ka$ in the parameters like number of states, free energy, pressure, entropy, chemical potential, internal energy, equation of state and specific heat. We use the following notation throughout the rest of the paper: a suffix `MS' is used to denote the corresponding quantity in the MS model that we have described so far. \\

\noindent \textbf{Number of States:} \\

Let us consider a non-relativistic ideal gas contained in a volume $ V = \int d^3 q $, $q$ being the generalized coordinates. Then the total number of permissible microstates $ \sum $ for the gas to occupy can be computed as \cite{pathria, greiner}
\be \sum = \int d^3 q ~ d^3 p = \frac{4 \pi V}{h^3} \int_0^{\infty} p^2 dp, \label{states} \ee which actually diverges as the energy or momentum has no finite upper bounds.
In contrast, for the MS model considered in this work, the number of available microstates to the system is given by
\be \sum_{MS} = \int d^3 q ~ d^3 p = \frac{4 \pi V}{h^3} \int_0^{\ka} p^2 dp, \label{statesms} \ee obviously a finite quantity due to the presence of the energy upper bound. This signifies the fact that in the MS model, the number of microstates is restricted to a much lesser value than its counterpart in normal relativistic or non-relativistic scenarios. \\

\noindent \textbf{Free Energy:} \\

Using the Stirling's approximation $ \ln [N!] \approx N \ln[N] - N $, the expression for free energy in the MS model $F_{MS}$ can be obtained as
\be 
\begin{aligned}
F_{MS} & = - k_B T ~ \ln [Z_{MS}(T, V, N)] \\
 & = - N k_B T \left[1 + \ln \left\{\frac{V}{N} \left[\left(\frac{2 \pi m k_B T}{h^2} \right)^{3/2} Erf \left(\frac{\ka}{\sqrt{2 m k_B T}}\right) - \frac{4 \pi m k_B T \ka}{h^3} e^{- \frac{\ka^2}{2 m k_B T}}\right]\right\}\right].  
\end{aligned} \label{freems} \ee
It can be immediately verified that in the limit $ \ka \rightarrow \infty $, the expression for free energy in MS model (\ref{freems}) reduces to the usual expression for free energy $F$ of a non-relativistic ideal gas given by
\be F = - N k_B T \left[1 + \ln \left\{\frac{V}{N} \left(\frac{2 \pi m k_B T}{h^2} \right)^{3/2} \right\}\right]. \label{free} \ee \\

\noindent \textbf{Pressure:} \\

For a non-relativistic ideal gas in usual case, its pressure $P$ can be calculated from (\ref{free}) using the relation
\be P = - \left(\frac{\partial F}{\partial V}\right)_{T, N} = \frac{N k_B T}{V}. \label{pressure} \ee
In the MS model considered in this paper, the pressure of non-relativistic ideal gas can be easily computed from (\ref{freems}) as
\be P_{MS} = - \left(\frac{\partial F_{MS}}{\partial V}\right)_{T, N} = \frac{N k_B T}{V}, \label{pressurems} \ee producing the same relation as in the usual scenario (\ref{pressure}). \\

\noindent \textbf{Entropy:} \\

Entropy is one of the most important thermodynamic parameter which is a measure of disorder in any system. Using the expression for free energy (\ref{freems}), we evaluate the entropy $S_{MS}$ of the ideal gas in MS model as \\
\be 
\begin{aligned}
S_{MS} & = - \left(\frac{\partial F_{MS}}{\partial T}\right)_{V, N} \\
& = N k_B \left[\frac{5}{2} + \ln \left\{\frac{V}{N} \left(\frac{2 \pi m k_B T}{h^2} \right)^{3/2} Erf \left(\frac{\ka}{\sqrt{2 m k_B T}}\right)- \frac{4 \pi m k_B T \ka}{h^3} e^{- \frac{\ka^2}{2 m k_B T}}\right\} \right] \\
& + N k_B \left[\frac{2 \pi \ka^3 e^{- \frac{\ka^2}{2 m k_B T}}}{4 \pi m k_B T \ka  e^{- \frac{\ka^2}{2 m k_B T}} - (2 \pi m k_B T)^{3/2} Erf \left(\frac{\ka}{\sqrt{2 m k_B T}}\right)} \right].
\end{aligned} \label{entropyms} \ee
In comparison, the entropy $S$ for a non-relativistic ideal gas in usual scenario is obtained from (\ref{free})
\be S = - \left(\frac{\partial F}{\partial T}\right)_{V, N} = N k_B \left[\frac{5}{2} + \ln \left\{\frac{V}{N} \left(\frac{2 \pi m k_B T}{h^2} \right)^{3/2} \right\} \right]. \label{entropy} \ee
One can readily check that (\ref{entropy}) can be obtained from (\ref{entropyms}) in the limit $ \ka \rightarrow \infty $. \\
In the following Figure 2, we have plotted side by side the entropy against temperature for the usual case and for the MS model where an invariant energy upper bound is present. It can be clearly seen from the figure that the entropy for the MS model increases at a much slower rate after reaching a finite temperature and finally it saturates to a finite constant value. For the normal case, the entropy continuously increases at a more rapid rate and ultimately diverges. As entropy is a direct measure of the number of microstates available to a system, this plot shows that the number of states is finite for an ideal gas in MS model. This is indeed true due to the presence of the energy upper bound $\ka$ in the theory.
\begin{figure}[htb]
	{\centerline{\includegraphics[width=10cm, height=6cm] {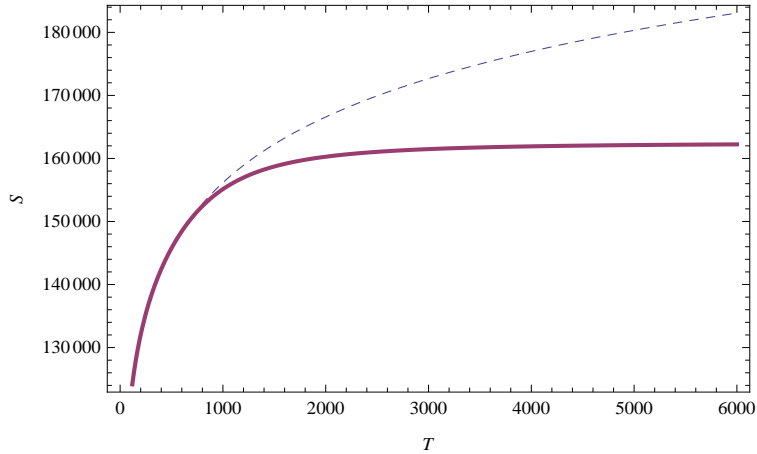}}}
	\caption{{\it{Plot of entropy of non-relativistic ideal gas $S$ against temperature $T$ for both in the usual theory and in MS model with an invariant energy scale. The dashed line corresponds to the usual result and the thick line represents the corresponding quantity in MS model considered in our work.}}} \label{fig2}
\end{figure} 

\newpage

\noindent \textbf{Chemical Potential:} \\

For the normal non-relativistic ideal gas, the chemical potential $ \mu $ is evaluated to be
\be \mu = \left(\frac{\partial F}{\partial N}\right)_{T, V} = - k_B T \ln \left[\frac{V}{N} \left(\frac{2 \pi m k_B T}{h^2} \right)^{3/2} \right]. \label{mu} \ee
In case of the MS model, the corresponding chemical potential is given by
\be \begin{aligned} \mu_{MS} & = \left(\frac{\partial F_{MS}}{\partial N}\right)_{T, V} \\
& = - k_B T \ln \left[\frac{V}{N} \left\{\left(\frac{2 \pi m k_B T}{h^2} \right)^{3/2} Erf \left(\frac{\ka}{\sqrt{2 m k_B T}}\right) - \frac{4 \pi m k_B T \ka}{h^3} e^{- \frac{\ka^2}{2 m k_B T}}\right\}\right]. 
\end{aligned} \label{mums} \ee
Consistent with the previous cases, we get back (\ref{mu}) from (\ref{mums}) in the limit $ \ka \rightarrow \infty $. \\

\noindent \textbf{Internal Energy:} \\

Once we have the expressions for free energy (\ref{freems}) and entropy (\ref{entropyms}) of non-relativistic ideal gas in MS model, it is straightforward to calculate the internal energy $U_{MS}$ as the following
\be U_{MS} = F_{MS} + T~S_{MS} = N k_B T \left[\frac{3}{2} + \frac{2 \pi \ka^3 e^{- \frac{\ka^2}{2 m k_B T}}}{4 \pi m k_B T \ka e^{- \frac{\ka^2}{2 m k_B T}} - (2 \pi m k_B T)^{3/2} Erf \left(\frac{\ka}{\sqrt{2 m k_B T}}\right)} \right]. \label{intms} \ee
It is quite obvious to see that in the $ \ka \rightarrow \infty $ limit, this expression (\ref{intms}) simplifies to the form of the internal energy of a normal non-relativistic ideal gas given by
\be U = F + T S = \frac{3}{2} N k_B T. \label{int} \ee

In the following Figure 3, we compare the internal energies for both the cases by plotting (\ref{int}) and (\ref{intms}) against temperature $T$. For the normal case, the plot is a straight line while the curve of internal energy asymptotically saturates to a finite value for the MS model. \\

\begin{figure}[htb]
	{\centerline{\includegraphics[width=10cm, height=6cm] {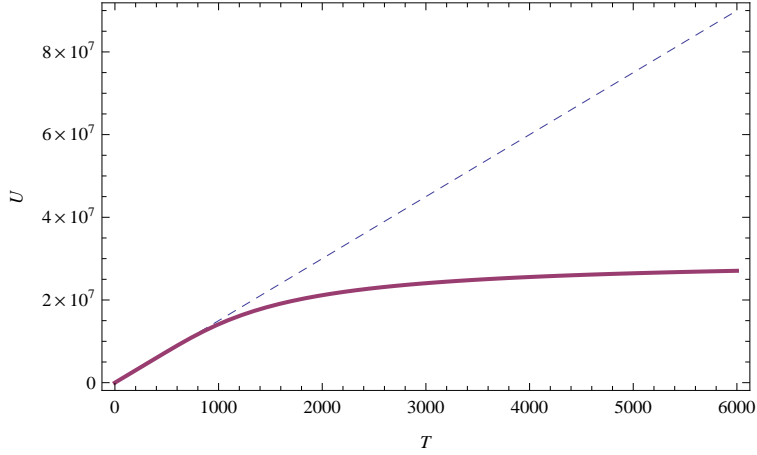}}}
	\caption{{\it{Plot of internal energy of non-relativistic ideal gas $U$ against temperature $T$ for both in the usual theory and in MS model with an invariant energy scale. The dashed line corresponds to the usual result and the thick line represents the corresponding quantity in MS model considered in our work.}}} \label{fig3}
\end{figure} 

\noindent \textbf{Equation of State:} \\

The energy density of a thermodynamic system is a measure of its internal energy $U$ per unit volume $V$. Thus, for a normal non-relativistic ideal gas, the energy density $\rho$ is obtained using (\ref{int}) as
\be \rho = \frac{U}{V} = \frac{3}{2} \frac{N k_B T}{V}. \label{rho} \ee
As we know the pressure $P$ of the ideal gas from (\ref{pressure}), the relation between energy density $\rho$ and pressure $P$, known as the Equation of State (EoS), for an ideal non-relativistic gas is easily found to be
\be P = \frac{2}{3} \rho. \label{eos} \ee
For the MS model, though the pressure $P_{MS}$ in (\ref{pressurems}) is the same as in (\ref{pressure}), but the internal energy $U_{MS}$ in (\ref{intms}) is quite different from (\ref{int}). Thus, using (\ref{pressurems}) and (\ref{intms}) and after some algebraic manipulation, we obtain the required Equation of State for a non-relativistic ideal gas in this MS model as
\be P_{MS} = \left[\frac{2}{3} - \frac{8 \pi \ka^3 e^{- \frac{\ka^2}{2 m k_B T}}}{36 \pi m k_B T \ka e^{- \frac{\ka^2}{2 m k_B T}} - 9(2 \pi m k_B T)^{3/2} Erf \left(\frac{\ka}{\sqrt{2 m k_B T}}\right) + 12 \pi \ka^3 e^{- \frac{\ka^2}{2 m k_B T}}}\right] \rho_{MS}.
\label{eosms} \ee
As $ \ka \rightarrow \infty $, the EoS for the MS model (\ref{eosms}) reduces to the usual EoS $P = \frac{2}{3} \rho$ for non-relativistic ideal gas as in (\ref{eos}). \\

\noindent \textbf{Specific Heat:} \\

Specific heat for constant volume of a non-relativistic ideal gas in MS model is obtained from (\ref{intms}) as
\be 
\begin{aligned}
C_{V, MS} & = \left(\frac{\partial U_{MS}}{\partial T}\right)_{V} \\
& = N k_B \left[\frac{3}{2} - \frac{\ka^3 \left(2 \ka \sqrt{m k_B T} e^{- \frac{\ka^2}{2 m k_B T}} + \sqrt{2 \pi} (\ka^2 - m k_B T) Erf \left(\frac{\ka}{\sqrt{2 m k_B T}}\right) \right)}{2 (m k_B T)^{3/2} \left(\sqrt{2 \pi m k_B T} Erf \left(\frac{\ka}{\sqrt{2 m k_B T}}\right) - 2 \ka e^{- \frac{\ka^2}{2 m k_B T}} \right)^2} \right].
\end{aligned} \label{cvms} \ee
For the conventional model, using (\ref{int}) the expression for specific heat $C_V$ of non-relativistic ideal gas is found to be 
\be C_V = \left(\frac{\partial U}{\partial T}\right)_{V} = \frac{3}{2} N k_B. \label{cv} \ee
It is easy to verify that the simple linear relation (\ref{cv}) can be obtained from (\ref{cvms}) when $ \ka \rightarrow \infty $. \\

\section{Discussion and Conclusion}

In this work, we have considered a particular model of Doubly Special Relativity (DSR), known as the Magueijo-Smolin (MS) model, in which an invariant upper bound of energy is present. We have thoroughly studied the implications of this energy upper bound $\ka$ in case of thermodynamics for a non-relativistic ideal gas. Starting with the modified dispersion relation in MS model, we have derived the expression for partition function of the ideal gas. This is one of the prime result of our work. Using this partition function, all the other thermodynamic parameters of an ideal gas for this MS model have been explicitly evaluated and their deviations from the usual non-relativistic case have been compared. It is clearly shown from our results that the number of microstates of the system is much lower in MS model than in the usual case. This is a very important result and is demonstrated analytically and graphically through simultaneous plots of the entropy for both the MS model and the normal case. This result also indicates towards a solution of the ``soccer-ball problem" for multi-particle systems in DSR \cite{maghoss}. All the results obtained in our work is consistent in the sense that they give back the conventional expressions in the limit $\ka \rightarrow \infty$. \\
As discussed earlier, we have confined this work to study the thermodynamic properties of an ideal gas for MS model in non-relativistic framework. A numerical study of the same considering the relativistic regime can be done in future. Also, thermodynamics of Fermion gas with this MS dispersion relation is another interesting area for future studies in which some modifications in the Fermi energy level can have astrophysical implications such as ``Chandrashekhar Mass Limit". Our another aim in future is to study the cosmological implications of this finite energy scale $\ka$ as it may help to circumvent the curvature singularity like the ``bouncing" loop quantum cosmology theories \cite{bojowald}. This in turn can induce a situation where the issue of big bang singularity can be avoided.

% ----------------------------------------------------------------

\bibliographystyle{amsplain}

\end{document}